\documentclass[conference,letterpaper]{IEEEtran}
\usepackage{bbm}
\usepackage{lineno}
\usepackage{amssymb}
\usepackage{amsthm}
\usepackage{epsfig}
\usepackage{graphicx}
\usepackage{graphics}
\usepackage{float}
\usepackage{subfigure}
\usepackage{multirow}
\usepackage{color}
\usepackage{comment}
\usepackage{makeidx}
\usepackage{xspace}
\usepackage{wrapfig}
\usepackage{lipsum}
\usepackage{mathtools}
\usepackage{cuted}
\usepackage{cite}
\usepackage{amsfonts}
\usepackage{textcomp}
\usepackage{enumerate}

\usepackage{textcomp}
\usepackage{enumerate}
\usepackage{algpseudocode}
\usepackage{algorithm}
\usepackage[dvipsnames]{xcolor}

\makeindex
\theoremstyle{plain}

\newtheorem{Theorem}{Theorem}
\newtheorem{Proposition}{Proposition}
\theoremstyle{definition}

\newtheorem{Lemma}{Lemma}

\newtheorem{Proof of Lemma}{Proof of Lemma}

\makeatletter
\renewcommand\@endtheorem{\vvv@endmarker\endtrivlist\@endpefalse}
\newcommand\vvv@endmarker{%
  {\unskip\nobreak\hfil\penalty50
  \hskip2em\vadjust{}\nobreak\hfil\openbox
  \parfillskip=0pt \finalhyphendemerits=0 \par
  \penalty 10000 \parskip=0pt\noindent}\ignorespaces}
\makeatother

\definecolor{darkred}{rgb}{1, 0.1, 0.3}
\definecolor{darkblue}{rgb}{0.1, 0.1, 1}
\definecolor{darkgreen}{rgb}{0,0.6,0.5}

\def \S {\mathcal{S}}

\def \T {\mathcal{T}}

\def \W {\mathcal{W}}

\def \I {\mathcal{I}}
\def \J {\mathcal{J}}
\def \F {\mathbb{F}}

\allowdisplaybreaks

\cleardoublepage
\begin{document}
\title{Analog Multi-User Linearly-Separable Distributed Computing Via Compressed Sensing}
\title{Multi-User Linearly-Separable Distributed Computing Via Compressed Sensing}
\title{Distributed Computing Meets Compressed Sensing}
\title{Multi-User Distributed Computing\\ Via Compressed Sensing}
  \author{%
  \IEEEauthorblockN{Ali Khalesi, Sajad Daei, Marios Kountouris, Petros Elia}
  \IEEEauthorblockA{Communication Systems Department, EURECOM\\
                    06410 Sophia Antipolis, France\\
                    Email: \{ali.khalesi;sajad.daei;marios.kountouris;petros.elia\}@eurecom.fr}\
}
  \maketitle
\begingroup\renewcommand\thefootnote{}
\footnotetext{This work was supported by the European Research Council (ERC) through
the EU Horizon 2020 Research and Innovation Program under Grant 725929
(Project DUALITY) and Grant 101003431 (Project SONATA)}
    \author{\IEEEauthorblockN{Ali Khalesi, Sajad Daei, Marios Kountouris, Petros Elia}}
    \maketitle
\maketitle
\begin{abstract}
The multi-user linearly-separable distributed computing problem is considered here, in which $N$ servers help to compute the real-valued functions requested by $K$ users, where each function can be written as a linear combination of up to $L$ (generally non-linear) subfunctions. Each server
computes a fraction $\gamma$ of the subfunctions, then communicates a function of its computed outputs to some of the users, and then each user collects its received data to recover its desired function. Our goal is to bound the  ratio between  the computation workload done by all servers  over the number of datasets.

To this end, we here reformulate the real-valued distributed computing problem into a matrix factorization problem and then into a basic sparse recovery problem, where sparsity implies computational savings. Building on this, we first give a simple probabilistic scheme for subfunction assignment, which allows us to upper bound the optimal normalized computation cost as $\gamma \leq \frac{K}{N}$ that a generally intractable $\ell_0$-minimization would give. To bypass the intractability of such optimal scheme, we show that if these optimal schemes enjoy $\gamma \leq - r\frac{K}{N}W^{-1}_{-1}(- \frac{2K}{e N r} )$ (where $W_{-1}(\cdot)$ is the Lambert function and $r$ calibrates the communication between servers and users), then they can actually be derived using a tractable Basis Pursuit $\ell_1$-minimization. This newly-revealed connection 
opens up the possibility of designing practical distributed computing algorithms by employing tools  and methods from compressed sensing. 

\end{abstract}


\section{Introduction}
Distributed computing plays an important role in speeding up non-linear and computationally hard computing tasks. As the complexity of these tasks increases, there is an ever rising need for novel parallel processing techniques that efficiently offload computations to groups of distributed servers. This same complexity increase also brings about many challenges, including, to name a few, computing accuracy~\cite{jahani2021codedsketch,wang2021price}, scalability~\cite{li2017scalable,haddadpour2019trading,soleymani2021analog}, privacy and security~\cite{sun2018capacity,soleymani2020distributed,khalesi2021capacity,soleymani2020privacy,soleymani2021list}, and latency and straggler mitigation~\cite{raviv2020gradient,lee2017speeding,egger2022efficient,kai1,behrouzi2020efficient}. There is also substantial research work on various explorations of communication vs. computation trade-off~\cite{kai1},\cite{kai2}, \cite{Parrinello1},\cite{Mohammad1}. Moreover, motivated by the applicability of deriving schemes that work on real numbers, {other interesting related settings, such as the Lagrange-coded secret sharing over real numbers~\cite{Katrine}, coded distributed polynomial evaluation over complex matrices~\cite{Soleymani10}, \cite{Soleymani20}, as well as the setting of secure distributed multiplication of real or complex matrices~\cite{Hollanti}, have emerged. For a detailed survey of related research works, the interested reader is referred to~\cite{ng2020survey,CIT-103}.}

This same aforementioned complexity increase, has also brought about various related frameworks such as MapReduce \cite{dean2008mapreduce} and Spark \cite{zaharia2010spark} that apply to broad classes of functions. Focusing on functions over finite fields, the recent work in \cite{Khalesi-2} proposed the so-called \emph{Multi-User Linearly-Separable Distributed-Computing framework}, which allows for distributed computation of functions that adhere to the very broad linearly separable format, which in turn captures various classes of linear and non-linear functions of practical interest\footnote{For more information on this, please see~\cite{Mallick1, Mallick2}.}. Such functions have the form
\begin{align}
    F(D_1,D_2,\hdots,D_L) = \sum\nolimits^{L}_{l=1}f_{l}g_{l}(D_{l}) \nonumber
\end{align} 
where $D_1,D_2,\hdots,D_L$ are the $L$ input datasets,$W_{l} = g_{l}(D_{l})$ are the computed outputs of basis subfunctions $g_{l}(D_{l})$, and $f_{l}$ are scalar coefficients.  
In the multi-user ($K$ users and $N$ servers) setting where each user asks for its own function, the work in~\cite{Khalesi-2} transformed the distributed computing problem into a simple (preferably sparse) matrix factorization problem over finite fields, and then proceeded to make the direct connection between distributed computing, matrix factorization, and a new coding theoretic approach. In particular, for a $K \times L$ \emph{demand matrix} $\mathbf{F}$ where each row describes the coefficients that define the function requested by a user, the problem was transformed into the factorization problem $\mathbf{F}=\mathbf{D}\mathbf{E}$, { where $\mathbf{D}$ and $\mathbf{E}$ are the decoding and encoding matrices respectively}. $\mathbf{E}$ dictated which server should compute which subfunction and then how each server should combine the computed outputs before transmitting, while the $K \times N$ decoding matrix $\mathbf{D}$ dictated which user should each server communicate to and how each user should combine the various received signals. Then, a solution was proposed that derived from the powerful class of covering codes, and from a new class of so-called \emph{partial covering codes}. In particular, the parity-check matrix from such codes played the role of $\mathbf{D}$, while then, after considering the columns of $\mathbf{F}$ as syndromes, the columns of $\mathbf{E}$ were identified as the coset leaders with minimum weight, thus guaranteeing the sparsest $\mathbf{E}$ and thus the least computational cost. For example, when $\mathbf{D}$ is derived from the basic class of covering codes over a $q$-ary field, then the corresponding normalized computational cost $\gamma \in [0,1]$ --- describing the fraction of all subfunctions each server had to compute --- took the form $\gamma = H_q^{-1}(K/N)$ where $H_q^{-1}(\cdot)$ is the functional inverse of the entropy function.

We are here though interested in computing functions directly over the reals, which will indeed constitute a substantial deviation from the finite field case. This emphasis on the real (or complex) domain is necessitated by the fact that computing a real-valued problem over a finite field (after discretization) may not be as practical as computing it directly over the reals, mainly because discretization may entail large precision costs and accuracy losses, as well as because finite field computations are notoriously slower than floating point operations. For that, we will consider real-valued functions over $L$ real-valued datasets (or equivalently, with $L$ component/basis subfunctions), and $N$ computing servers and $K$ users each demanding their own function. 
As we are now working in the field of real numbers, the coding theoretic approach in~\cite{Khalesi-2} does not directly apply, and thus a new approach is required. 

Here, our approach is based on establishing, for the first time, a connection between distributed computing and compressed sensing. As a first step, we show (Proposition \ref{C}) that there exists an achievable scheme whose normalized computational cost is bounded above as $\gamma\leq \frac{K}{N}$. This is a probabilistic scheme, where $\mathbf{D}$ is chosen from the Gaussian ensemble, and where the corresponding sparsity of $\mathbf{E}$ is the outcome of a randomized process. Then we propose $\ell_{0}$-minimization, which takes as input $\mathbf{D}$ and $\mathbf{F}$ to yield a sparse $\mathbf{E}$. This minimization though is generally intractable, and for this reason, we draw from the rich literature of compressed sensing to suggest a more practical approach where we show (Theorem~\ref{CC}) that as long as there exists a scheme whose computational cost is bounded by $\gamma \leq - r\frac{K}{N}W^{-1}_{-1}(- \frac{2K}{e N r} )$ (where $W_{-1}(\cdot)$ is the Lambert function and $r$ is a parameter that calibrates the communication between servers and users) we can in fact employ a tractable basis pursuit $\ell_1$-minimization to derive such scheme.

\noindent{\bf Notations:}
For matrices $\mathbf{A}$ and $\mathbf{B}$, then $[\mathbf{A},\mathbf{B}]$ indicates the horizontal concatenation of the two matrices. We define $[n] \triangleq \{1,2,\hdots , n\}$. For any matrix $\mathbf{X} \in \F^{m \times n}$, then $\mathbf{X}(i,j),\: i \in [m],\: j \in [n]$, represents the entry in the $i$-th row and $j$-th column, while $\mathbf{X}(i,:),\: i \in [m]$, represents the $i$-th row, and $\mathbf{X}(:,j),\: j \in [n]$ represents the $j$-th column of $\mathbf{X}$. For two index sets $\I\subset [m], \J\in [n]$, then $\mathbf{X}(\I,\J)$ represents the sub-matrix comprised of the rows in $\I$ and columns in $\J$. We will use $\left\lVert \mathbf{X}\right\lVert_{0}$ to represent the number of nonzero elements of some matrix (or vector) $\mathbf{x}$. Also, $\otimes $ is the Kronecker product and $\text{vec}(\mathbf{X})$ is the vectorization of $\mathbf{X}$. 
\section{System Model and Problem Formulation}\label{System-Model}
We consider the multi-user linearly-separable distributed computation setting (cf.~Fig.~\ref{Fig: System Model}), which consists of $K$ users/clients, $N$ active servers, and a master node that coordinates servers and users. The tasks performed on each server may entail substantial computational complexity as well as time constraints. We consider a setting where each server $n$ can communicate in a single shot (a single time-slot) to some arbitrary user-set $\T_{n} \subset [K]$, via a dedicated broadcast channel.

In our setting, each user asks for a (generally non-linear) function from a space of linearly-separable functions, where each such function takes several datasets $D_1,\dots,D_L$ as input, and is of the form of a linear combination of individual subfunctions $g_{l}(D_l) \in \mathbb{R}$, each taking a single dataset $D_l$ as input. Thus, the function $F_k(D_1,\dots,D_L) \in \mathbb{R}$, demanded by user $k \in[K]$, is a real-valued function of the form
\begin{align}
    F_k(D_1,D_2,\hdots,D_L)&\triangleq f_{k,1}g_{1}(D_1) \hdots +f_{k,L}g_{L}(D_L)\:\:\\
    &=f_{k,1}W_1  +\hdots +f_{k,L}W_L\:\:\label{DefinitionOfLSFunctions}
\end{align}
where $W_l = g_{l}(D_l) \in \mathbb{R},\: l \in [L]$ is a so-called `file' output, and  $f_{k,l} \in \mathbb{R} ,\: k \in [K], l \in [L]$ are the linear combination coefficients that define each desired function.

\subsection{Phases of the Process}
The model involves three phases, with the first being the \emph{demand phase}, the second being the \emph{assignment and computation phase}, and the final one the \emph{transmission and decoding phase}. In the {demand phase}, each user $k \in [K]$ requests $F_k(\cdot)$ from the master node, who then deduces the decomposition as in~\eqref{DefinitionOfLSFunctions}.
Then, based on these $K$ desired functions, during the assignment and computation phase, the master assigns some of the subfunctions to each server $n$, which then proceeds to compute these and produce the corresponding files $W_l = f_{l}(D_l)$ for all the subfunctions $f_{l}(D_l), l \in \W_{n}$ it is responsible for. 

During the transmission phase, each server $n \in [N]$ broadcasts 
\begin{align}
  z_{n}\triangleq \sum_{l \in [L]} e_{n,l} W_l,\:\: n\in [N] \label{EncodedFiles}
\end{align}
in a single shot its own linear combination of the locally computed output files, and does so to its own particular subset of users $\T_{n}$. The above is defined by the so-called \emph{encoding coefficients} $e_{n,l}\in \mathbb{R}$ which are determined by the master.

Finally, during the decoding phase, each user $k$ linearly combines the received signals as follows
\begin{align}
    F'_{k} \triangleq \sum_{n \in [N]} d_{k,n} z_{n}\label{DecedFiles}
\end{align}
for some decoding coefficients $d_{k,n} \in \mathbb{R}, n\in [N]$, determined again by the master node. Naturally $d_{k,n} =0,\forall k \notin \mathcal{T}_{n}$. 
In the end, 
 we say the exact decoding is successful when  $F'_k = F_k$ for all $k\in[K]$.
\begin{figure}
\centering
\includegraphics[scale=0.5]{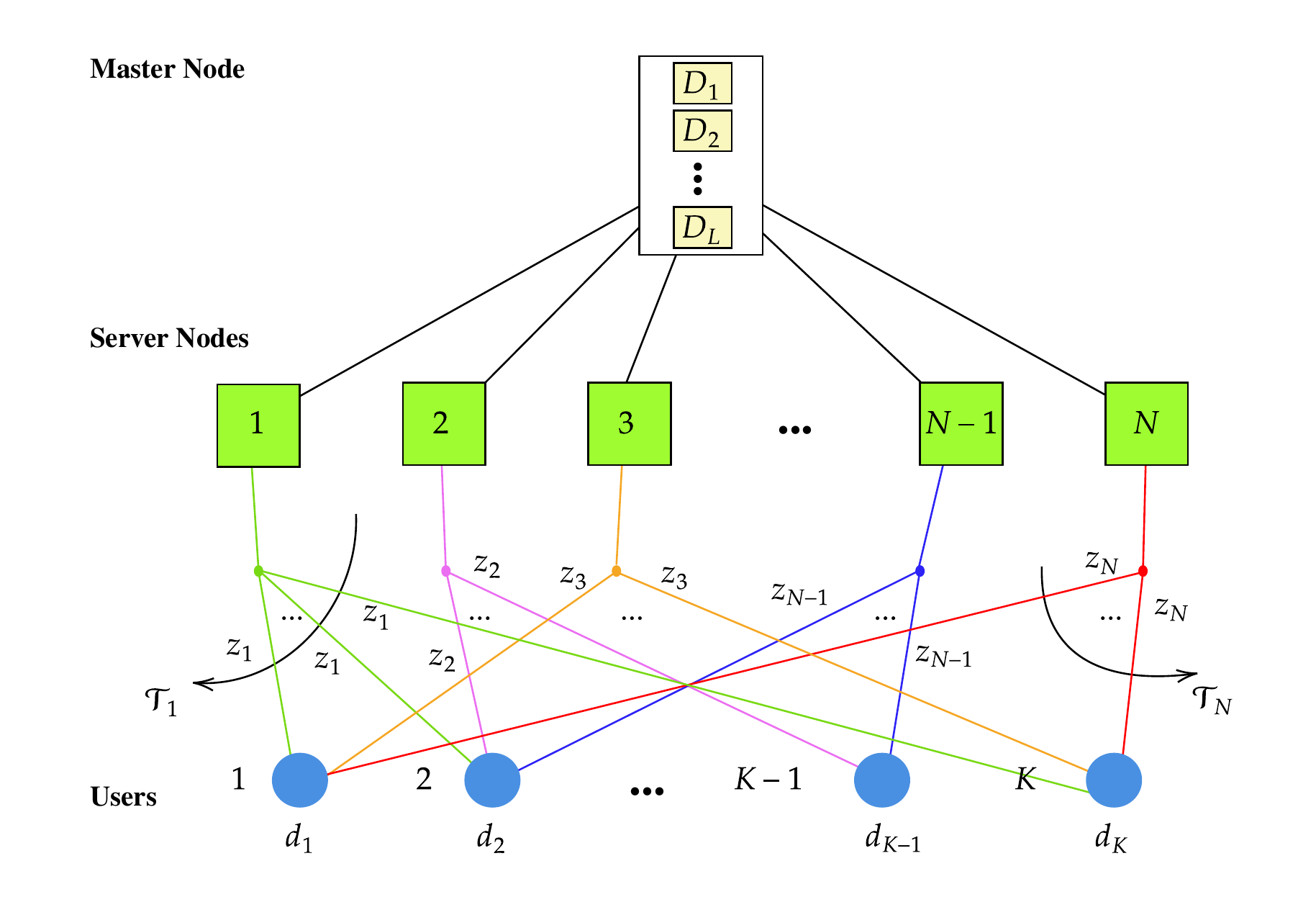}
\caption{The $K$-user, $N$-server, $L$-Dataset linearly-separable computation setting. Once each user informs the master of its desired function $F_k(\cdot)$, each server $n \in [N]$ computes a  subfunction $W_l = f_{l}(D_l) \in \mathbb{R}$  in $\W_n \subseteq [L]$. Afterwards, server $n$ broadcasts a linear combination ${z}_{n}$ (of the locally available computed files) to all users in $\T_{n}$. This combination is defined by the coefficients $e_{n,l}$. Finally, for decoding, each user $k \in [K]$ linearly combines (based on decoding vectors $\mathbf{d}_k$) all the received signals from all of the servers it has received from. Decoding should produce for each user its desired function $F_k(D_1,\dots,D_L)$.}
\label{Fig: System Model}
\end{figure}
\subsection{Problem Formulation}\label{Formulating}
Similarly to the finite field case, also here, to formulate the problem we use $\mathbf{f}\triangleq [F_1,F_2,\hdots,F_K]^{\intercal},\:  \mathbf{f}_k \triangleq [f_{k,1},f_{k,2},\hdots,f_{k,L}]^{\intercal},\: k \in [K]\label{function-vectors-1},
    \mathbf{w}\triangleq [W_{1},W_{2},\hdots,W_{L}]^{\intercal}$
where $\mathbf{f}$ represents the vector of the demanded functions outputs (cf.~\eqref{DefinitionOfLSFunctions}), $\mathbf{f}_k$ the vector of function coefficients for user $k$ (cf.~\eqref{DefinitionOfLSFunctions}), and $\mathbf{w}$ the vector of output files combined over all subfunctions. 
We also have $\mathbf{e}_{n} \triangleq [e_{n,1},e_{n,2},\hdots, e_{n,L}]^\intercal,\: n \in [N]
     \mathbf{z} \triangleq [z_{1}, z_{2},\hdots, z_{N}]^ \intercal$
respectively representing the encoding vector at server $n$, and the overall transmitted vector across all the servers (cf.~\eqref{EncodedFiles}).
Furthermore, we have $ \mathbf{d}_{k} \triangleq [d_{k,1},d_{k,2},\hdots, d_{k,N}]^ \intercal,\: k \in [K]
      \mathbf{f}'\triangleq [F'_1,F'_2,\hdots,F'_K]^{\intercal}$

respectively representing the decoding vector at user $k$, and the vector of the decoded functions across all the users.
In addition, we have $ \mathbf{F} \triangleq [\mathbf{f}_1,\mathbf{f}_2,\hdots,\mathbf{f}_K]^{\intercal} \in \mathbb{R}^{K \times L}, 
        \mathbf{E} \triangleq [\mathbf{e}_{1},\mathbf{e}_{2},\hdots, \mathbf{e}_{N}]^\intercal \in \mathbb{R}^{N \times L},
    \mathbf{D} \triangleq [\mathbf{d}_1,\mathbf{d}_2, \hdots , \mathbf{d}_K]^{\intercal} \in \mathbb{R}^{K \times N}$ 

where $\mathbf{F}$ represents the $K\times L$ so-called \emph{jobs matrix} of all function coefficients across all the users, where $\mathbf{E}$ represents the $N\times L$ \emph{computing and encoding matrix} across all servers, and where $\mathbf{D}$ represents the $K\times N$ \emph{decoding matrix} across all the users.

Directly from~\eqref{DefinitionOfLSFunctions}, we have that
\begin{align}
    \mathbf{f} =[\mathbf{f}_1,\mathbf{f}_2,\hdots,\mathbf{f}_K]^{\intercal} \mathbf{w}\label{Functions-one}
\end{align}
and from \eqref{EncodedFiles} we have the overall transmitted vector taking the form
\begin{align}
    \mathbf{z} =[\mathbf{e}_{1}, \mathbf{e}_{2}, \hdots,\mathbf{e}_{N}]^\intercal \mathbf{w} = \mathbf{E} \mathbf{w}.  \label{EncodedCashedData-1}
\end{align}
Furthermore, directly from~\eqref{DecedFiles} we have that
\begin{align}
    F'_k= \mathbf{d}_{k}^{T} \mathbf{z}
\end{align}
and thus we have
\begin{align}
    \mathbf{f}' = [\mathbf{d}_1,\mathbf{d}_2,\hdots,\mathbf{d}_K]^{\intercal} \mathbf{z} = \mathbf{D}\mathbf{z}.\label{DecodedData-one}
\end{align}
Recall that we must guarantee 
\begin{align}
    \mathbf{f}'=\mathbf{f}.\label{feasibility-one}
\end{align}
After substituting \eqref{Functions-one}, \eqref{EncodedCashedData-1} and \eqref{DecodedData-one} into \eqref{feasibility-one}, we see that the above feasibility condition in~\eqref{feasibility-one} is satisfied if and only if
\begin{align}
    \mathbf{D}\mathbf{E}\mathbf{w} = \mathbf{F}\mathbf{w}.\label{MainEquationWithW}
\end{align}
For this to hold for any $\mathbf{w}$, we must thus have
\begin{align}
    \mathbf{D}\mathbf{E} = \mathbf{F}.\label{MainEquation}
\end{align}

\subsection{Computational Cost}
Recalling quickly that each server $n$ computes the subfunctions whose index are in $\W_{n}$, and since $\W_{n} = \mathrm{supp}(\mathbf{E}({n}, :))$, then the \emph{normalized computation cost} in our case naturally takes the form 
\begin{align}
    \gamma \triangleq \frac{\sum^{N}_{n =1} |\W_n|}{NL} = \frac{\left\lVert \mathbf{E} \right\lVert_{0}}{NL}.\label{normalized-computation}
\end{align}
As one can see, $\gamma$ simply describes the average fraction of subfunctions that must be computed by each server, which is also the fraction of non-zero elements in $\mathbf{E}$. It is now clear that decomposing $\mathbf{F}$ into the product of  two  matrices $\mathbf{D}$ and sparse $\mathbf{E}$, implies reduced computation cost which  results in  reduced delay. In particular, the fewer number of nonzero elements in  a row of $\mathbf{E}$ means less delay in finishing up a task  for a server.  

\section{Results}\label{Results}
In this section, we first give a basic probabilistic scheme for subtask assignment, based on 
employing a Gaussian\footnote{This implies that each entry of $\mathbf{D}$ is independently and identically picked from a Gaussian distribution.} random matrix $\mathbf{D}$, where the scheme employs a simple zero-forcing approach that solves a determined linear system. Albeit basic, this will allow us to upper bound the optimal normalized computation cost --- which a generally intractable $\ell_0$-minimization would give --- as $\gamma \leq \frac{K}{N}$.

\begin{Proposition}\label{C}
For the multi-user linearly-separable distributed computing problem, with $K$ users, $N$ servers and $L$ datasets, employing a random Gaussian $\mathbf{D}$, guarantees that with probability $1$, there exists a scheme with bounded normalized computation cost $\gamma \leq K/N$, which serves as an upper bound the $\ell_0$-minimal cost.
    \end{Proposition}
    \begin{proof}
        From \eqref{MainEquation}, we have that $ \mathbf{F}(:,l) =\mathbf{D}\mathbf{E}(:,l), \mathbf{F}(:,l) \in \mathbb{R}^{K \times 1},\: \forall l \in [L]$ where for each $l$, in the context of the $\ell_{0}$-minimization in~\eqref{opt-1}, we have $\mathbf{y}=\mathbf{F}(:,l)$, $\mathbf{D}=\mathbf{A}$ and $\mathbf{E}(:,l)=\mathbf{z}$, where again we have an underdetermined system of equations.

Consider choosing $L$ arbitrary random subsets $\mathcal{S}_l \subset [N], \ l \in [L]$ where $|\mathcal{S}_l| = K$. Now for each $l\in [L]$, we focus on the $l-{th}$ column $\mathbf{E}([N],l)$ of $\mathbf{E}$ and set the elements $\mathbf{E}([N] \backslash \mathcal{S}_l,l)=0$, i.e., from column $l$, only the elements indexed by $\mathcal{S}_l$ remain non-zero. Now since 
$\mathbf{F}(:,l)= \mathbf{D}([K],[N]\backslash \mathcal{S}_{l}) \mathbf{E}([N]\backslash \mathcal{S}_{l},l)+ \mathbf{D}([K], \mathcal{S}_{l})\mathbf{E}(\mathcal{S}_{l},l)$, we get 
$\mathbf{F}(:,l) = \mathbf{D}([K],\S_l) \mathbf{E}(\S_l,l)$, which is a determined system of equations that allows us to determine $\mathbf{E}(\S_l,l)$. 
In the above, $\mathbf{D}([K],\S_l)$ is the corresponding $K \times K$ submatrix  of $\mathbf{D}$. Given the above, each such $\mathbf{D}([K],\S_l)$ is a $K \times K$ Gaussian sub-matrix, which is naturally nonsingular with probability one. Thus, the determined system of equations always has a unique solution for $\mathbf{E}(\S_l,l), \forall l \in [L]$, therefore the scheme works for all $F(:,l),\: l \in [L]$.
This scheme guarantees that  $ \left\lVert \mathbf{E}(:,l)\right\lVert_{0} \leq K$, then $ \left\lVert \mathbf{E}\right\lVert_{0} \leq KL$, and thus guarantees that $\gamma_{} \leq \frac{K}{N}$. Better performance can be achieved by employing $\ell_0$-minimization as in~\eqref{opt-1}. 
     \end{proof}
As we will discuss in Section~\ref{CCSS2}, $\ell_0$-minimization is known to be NP-hard~\cite{foucart13}, hence intractable. To offer a practical solution, the following theorem utilizes results from compressed sensing to describe a range of practical solutions, which now use $\ell_{1}$-minimisation in order to find --- as we will clarify later on --- sparse (and unique) encoding matrices $\mathbf{E}$. 
\begin{Theorem} \label{CC}
For the multi-user linearly-separable distributed computing problem, with $K$ users, $N$ servers and $L$ datasets, if a scheme exists with a ($\kappa,\beta$) sub-Gaussian random matrix $\mathbf{D}$ (cf. Lemma~\ref{lemmaFast}) for which $\ell_{0}$-minimisation would yield
        \begin{align}
          \gamma_{} \leq - \frac{1}{r}\frac{K}{N} W^{-1}_{-1}(- \frac{2K}{e r N} ) ,\:\:\:\: 0\leq K/N \leq 12 (2 \beta + \kappa)/ \kappa^{2}, \nonumber
        \end{align}
then the corresponding (and unique) $\mathbf{E}$ can be found via basis pursuit $\ell_1$-minimization with probability at least  $1-2e^{-\frac{KL}{r}}$, where $r =12 (4 \beta + 2\kappa)/ \kappa^{2}$.
\end{Theorem}

\begin{proof}
The proof is provided in the following section, which starts with a brief primer on compressed sensing.    
\end{proof}

\section{Proof of Theorem~\ref{CC}\label{CCSS2}} 

Before proceeding with the proof, we quickly describe some basic properties of compressed sensing. 
\subsection{Brief Primer on Compressed sensing} 
We provide here a brief introduction of the compressed sensing results~\cite{foucart13,candes_RIPless,candes2006stable,candes2008restricted,candes2006near,cai2010new,Cai12,FOUCART201097,foucart2009sparsest,mo2011new,daei2019living}, which will be employed in our distributed computing problem.  We will utilize notation common to the compressed sensing literature, and the link to the computing parameters will be clarified in the next subsection. 

As described in~\cite{candes2006near}, compressed sensing seeks to recover a sparse vector $\mathbf{x}\in\mathbb{R}^{p}$ from a few underdetermined linear measurements of the form: 
\begin{align}
    \mathbf{y}=\mathbf{A x}\in\mathbb{R}^{m} \label{CS-eq1} 
    \end{align}
where $\mathbf{A}\in\mathbb{R}^{m\times p}, m,p \in \mathbb{N}$ is the so-called measurement matrix, and $\mathbf{y}=[y_1,..., y_m]^T$ is the measurement vector. In our case, as we will see later on, $\mathbf{y}$ will be associated to our computing and encoding matrix $\mathbf{E}$, then $\mathbf{A}$ to the communication and decoding matrix $\mathbf{D}$, and $\mathbf{x}$ will be associated to the jobs matrix $\mathbf{F}$. The general approach is to recover the sparsest solution via a basic but computationally intractable $\ell_0$-minimization that takes the form
\begin{align}
\min_{\mathbf{z}\in\mathbb{R}^{p}}  \|\mathbf{z}\|_0:=\sum_{i=1}^p 1_{|z_i|\neq 0} ~~\text{ subject  to} ~~  \mathbf{y} = \mathbf{A} \mathbf{z} \label{opt-1}
\end{align}
where $1_{|z_i|\neq 0}$ denotes the indicator function.  This same optimization will lead to the sparsest solution for $\mathbf{E}$ and thus will yield the smallest possible $\gamma$. To the best of our knowledge, there are no results that enables us to bound the weight of this sparsest solution, and for that we will use a basic constructive approach to bound $\gamma$.

The NP-hard nature of the optimization problem in~\eqref{opt-1} has led to the consideration of an $\ell_1$-norm minimization approach, also known as \emph{basis pursuit}, which is considered as the closest convex tractable alternative for \eqref{opt-1}, and which is given by
\begin{align}\label{eq.L1}
\min_{\mathbf{z}\in \mathbb{R}^{p}} &\|\mathbf{z}\|_1:=\sum_{i=1}^p |z_i|\\\
&{\rm s.t. }~~ \mathbf{y}=\mathbf{A z}.
\end{align}
It is well established in compressed sensing that the estimate $\widehat{\mathbf{x}}\in\mathbb{R}^{p}$ obtained by solving \eqref{eq.L1}, achieves the desired unique solution $\mathbf{x}$ as long as some conditions are satisfied\footnote{For our computing problem. these will be conditions in the form of an upper bound on $\gamma$.}. These conditions are closely related to certain properties of the measurement matrix, with one such condition being the well known restricted isometry property (RIP)~\cite{Candes1}, which dictates how well $\ell_1$-norm optimization algorithms, such as basis pursuit~\cite{chen2001atomic}, can perform. 
This is captured in the following result, found in~\cite[Theorem 6.2]{foucart13}, which is reproduced here.

\begin{Lemma}\label{RIP-bound}
For a matrix $\mathbf{A}  \in \mathbb{R}^{m \times p}$ and for  
\begin{align}
    \delta_s(\mathbf{A}) \triangleq \underset{\mathcal{S} \subset [N], |\mathcal{S}| \leq s}{\max} \left \lVert \mathbf{A}^{*}_{\mathcal{S}} \mathbf{A}_{\mathcal{S}} - \mathbf{I}\right\lVert^{2}_{2 \rightarrow{}2}
\end{align}
being the $s$th restricted isometry constant, and if 
$\delta_{2s}(\mathbf{A}) < \frac{1}{3}$, then every $s$-sparse vector $\mathbf{x} \in \mathbb{R}^{p}$ is the unique solution of 
\begin{align}
  \underset{\mathbf{z} \in \mathbb{R}^{p}} {\text{minimize}}  \left\lVert \mathbf{z} \right\lVert_{1} \:\:\: \text{subject to}\:\:\: \mathbf{A}\mathbf{z}= \mathbf{A}\mathbf{x}.
\end{align}
\end{Lemma}
{{In particular, the above Lemma  shows that having a measurement vector $\mathbf{y} \in \mathbb{R}^{m} $, where we know apriori that  it is a result of a linear system $\mathbf{A} \mathbf{x}, \: \mathbf{x} \in \mathbb{R}^{p}, m \leq p$, induced by a unique and $s$-sparse vector $\mathbf{x}$ if $\delta_{2s}(\mathbf{A}) < \frac{1}{3}$,
then via having  a $\ell_1$-minimizer, we can  find $\mathbf{z}$ as a solution to $\mathbf{A}\mathbf{z}$, which  it has  minimum $\left\lVert \mathbf{z} \right\lVert_{1}$}, then the above Lemma guarantees that the solution of this minimization is equal to $\mathbf{x}$. }
The above lemma shows that having $\delta_{2s}(\mathbf{A}) < 1/3$ is sufficient to guarantee the exact recovery of all unique $s$-sparse vectors via $\ell_1$-minimization. 
It basically states that if $\mathbf{A}$ behaves relatively similar to orthonormal matrices when operating on sparse vectors, then $\ell_1$-minimization will act as an $\ell_0$-minimization.

In our problem here, it is important that we pick $\mathbf{D}$ such that $\mathbf{A}$ abides by the above property. The following two results tell us directly how to do that. The first, below, is directly adapted from~\cite[Theorem 9.2]{foucart13}. 
\begin{Lemma}\label{lemmaFast}
Let $\mathbf{A}\in\mathbb{R}^{m\times p}$ be an i.i.d. (zero mean, unit variance) sub-Gaussian random matrix with parameters $\beta$ and $\kappa$ such that 
\begin{align}
    \mathbb{P}(|A_{i,j}|\ge t)\le \beta {\rm e}^{-\kappa t^2}, \forall t>0. \label{subgaussin-param}
\end{align}
Then, $\delta_{2s}(\frac{\mathbf{A}}{\sqrt{m}}) \le \delta$ is satisfied with probability at least $1-2 {\rm e}^{-\frac{\delta^2 m}{2c}}$ for any $\delta$ such that 
\begin{align}
    m\ge 2c \delta^{-2} s \ln (\frac{ep}{2s})
\end{align}
where 
\begin{align}
    c = \frac{2(4 \beta + 2k)}{3 k^2}.
\end{align}
\end{Lemma}
Combining Lemma~\ref{lemmaFast} and Lemma~\ref{RIP-bound} after setting $\delta=1/3$, implies that the uniform recovery of all $s$-sparse vectors is possible with high probability via the $\ell_1$-minimization in~\eqref{eq.L1} as long as the number of measurements satisfies $m \geq (12 (4 \beta + 2\kappa)/ \kappa^{2}) s \ln(\frac{e p}{2s})$ .

\subsection{Proof of Theorem \ref{CC}}\label{T-2}
Directly from~\eqref{MainEquation}, we have
\begin{align}
\text{vec}{(\mathbf{F}}) = (\mathbf{D} {\otimes}   I_{L \times L}) \times  \text{vec}{(\mathbf{E})},        \label{reformulating-1}
\end{align}
which matches the compressed sensing setting $\mathbf{y}=\mathbf{A x}$ in~\eqref{CS-eq1} when considering $\mathbf{y} = \text{vec}{(\mathbf{F})}$, $\mathbf{A} = \mathbf{D} {\otimes}   I_{L \times L}$, and $\mathbf{x} = \text{vec}{(\mathbf{E})}$, where now $m=KL$ and $p = NL$.

Furthermore, let us also note that directly from~\cite{jokar2009sparse}, we have
\begin{align}
    \delta_s(\mathbf{D}\otimes \mathbf{I}_{L\times L}) \leq \delta_s(\mathbf{D}). \label{RIP-Koronecker}
\end{align}

With the elements of $\mathbf{D}$ being chosen independently from a zero-mean, unit-variance sub-Gaussian distribution with parameters $\beta$, $\kappa$ (cf. \eqref{subgaussin-param}), we can now employ Lemma~\ref{RIP-bound} and~\eqref{RIP-Koronecker}, together with Lemma~\ref{lemmaFast} after setting $\delta_{2s}< \delta= 1/3$, to conclude that the exact recovery threshold for a unique $\mathbf{E}$ matrix via $\ell_1$-minimization driven by basis pursuit, takes the form 
      \begin{align}
          KL \geq r \left  \lVert \mathbf{E} \right\lVert_{0} \ln(\frac{e NL}{2 r\left\lVert \mathbf{E} \right\lVert_{0}})
      \end{align}
      where $r=12 (4 \beta + 2\kappa)/ \kappa^{2}$. Then after normalizing both sides by $NL$ and applying~\eqref{normalized-computation}, we get
     \begin{align}
          \frac{K}{N} \geq  r \gamma \ln(\frac{e}{2 r \gamma} ). \label{r1}
      \end{align}
      Let us now define $f(x) \triangleq  - r^{-1}x W^{-1}_{-1}(- \frac{2x}{e })$ and  evaluate it on the  both sides of \eqref{r1}, at $x_1=K/N$  and $x_2= r \gamma\ln(\frac{e}{2 r \gamma} )$ since $f(x)$ is a monotonically increasing function on  $0\leq x \leq r/2$ 
 and $ r/2\geq x_1\geq x_2 \geq 0$, we have $f(x_1) \geq f(x_2)$, in the view of the fact that its inverse\footnote{To see this, for $g(x) \triangleq  r x \ln(\frac{e}{2 r x})$ and $f(x) \triangleq -r^{-1} x W^{-1}_{-1}(-2x /e)$, we see that $g^{-1}(x)=f(x)$ simply because $
            f(g(x))=  - r^{-1}g(x) W^{-1}_{-1}(- \frac{2g(x)}{e })= -x \ln(\frac{e}{2 r x}) W^{-1}_{-1}= -x \ln(e/2rx)W^{-1}_{-1}(\frac{2rx}{e} \ln(2rx/e))=-x\ln(e/{2rx})/\ln(2rx/e)=x$.} function is $f^{-1}(x)= r x\ln(\frac{e}{2r x})$, we can retrieve the claim. 
            
      \section{Discussion and Conclusion} \label{Discussion}
In the context of our distributed computing problem, it is interesting to observe some of the similarities that exist between the real case (which employed compressed sensing techniques) and the finite-field case in~\cite{Khalesi-2}, which employed the structure of covering codes whose covering radius was a measure of the sparsity of the solution for $\mathbf{E}$. For example, in the extreme case of $L=q^{K}$, the work in ~\cite{Khalesi-2} revealed\footnote{Over a $q$-ary alphabet, $ H_q(x)$ denotes the entropy function, which takes the form $ H_q(x) \triangleq  x \log_q(q-1) - x \log_q(x) - (1-x) \log_q(1-x)$ for all $0<x<1-1/q$.} the optimal normalized computational cost to be of the form of $\gamma \simeq H_q^{-1}(K/N)$, which almost matches $\gamma \simeq K/N$ in the limit of large $q$, derived in this paper.

It is also worth noting that our result in Theorem \ref{CC} automatically accepts an additional uniqueness property --- on the sparsest solution $\mathbf{x}$ in~\eqref{CS-eq1} --- which is in fact not needed in our distributed computing problem. It would be interesting to explore further improvements in the computational costs, upon the removal of this uniqueness condition. Finally, one can imagine that further improvements in the distributed computing problem could also benefit from the deep connections revealed in~\cite{candes2006near} between compressed sensing and error correction. 

\bibliographystyle{ieeetr}
\bibliography{ref}
\end{document}